# Construction of Optical Spatiotemporal Skyrmions

Houan Teng[1], Xin Liu[1,2,3], Nianjia Zhang[1], Haihao Fan[1], Guoliang Chen[1], Qian Cao[1,4], Jinzhan Zhong[1,4], Xinrui Lei[1,4,*], Qiwen Zhan[1,4,5,*]

[1]School of Optical-Electrical and Computer Engineering, University of Shanghai for Science and Technology, Shanghai 200093, China
[2]Shandong Provincial Engineering and Technical Center of Light Manipulations and Shandong Provincial Key Laboratory of Optics and Photonic Device, School of Physics and Electronics, Shandong Normal University, Jinan 250014, China.
[3]Collaborative Innovation Center of Light Manipulations and Applications, Shandong Normal University, Jinan 250358, China.
[4]Zhangjiang Laboratory, 100 Haike Road, Shanghai 201204, China
[5]International Institute for Sustainability with Knotted Chiral Meta Matter (WPI-SKCM2), Hiroshima University, Higashihiroshima, Hiroshima, 739-8526, Japan.

*Corresponding authors: xrlei@usst.edu.cn; qwzhan@usst.edu.cn

**Abstract:**
The creation and manipulation of photonic skyrmions provide a novel degree of freedom for light-matter interactions, optical communication and nanometrology. Since the localized vortex within skyrmions arises from the twist and curl of the phase structure, the orbital angular momentum of light is essential for their construction. While numerous skyrmionic textures have been proposed, they are formed within the spatial domain and induced by the longitudinal orbital angular momentum. Here we theoretically propose and experimentally observe spatiotemporal skyrmions within a picosecond pulse wavepacket, generated through vectorial sculpturing of spatiotemporal wavepackets. The skyrmionic textures emerge within the spatiotemporal distribution of a vector field encompass all possible polarization states. Constructed upon the transverse orbital angular momentum, spatiotemporal skyrmions, in contrast to spatial skyrmions, exhibit no helical twisting perpendicular to the skyrmion plane, demonstrating enhanced stability to deformations or perturbations. These results expand the skyrmion family and offer new insights into optical quasiparticles, potentially leading to advanced applications in optical metrology, sensing, and data storage.

## Introduction

Skyrmions are topologically stable quasiparticles characterized by nontrivial textures, which cannot be continuously deformed into a trivial state due to their inherent, indestructible topological twist. Initially proposed within the theoretical framework of elementary particles[1–3], skyrmions have subsequently emerged in diverse physical systems, including Bose-Einstein condensates[4–6], nematic liquid crystals[7–9], and chiral magnets[10,11]. The ultracompact size and inherent stability[12,13], attributed to topological protection[14,15], offer significant potential for applications in data storage and spintronic devices[16,17]. In recent years, skyrmionic textures have been introduced into the realm of optics[18–20], where topological structures are realized through the precise manipulation of electromagnetic fields across various synthetic dimensions. On one hand, the spin-orbit interaction of light in nonparaxial beams offers a versatile platform for sculpting electromagnetic fields[20–22]. This dynamic mutual conversion between angular momentum components facilitates the formation of spin or field skyrmions[20,23–25]. On the other hand, Stokes skyrmions have been generated in paraxial beams[26–37], where the topological features are encoded within the polarization texture of structured light and manipulated through Poincaré beams[38]. The robustness of photonic skyrmions, demonstrated through their resilience to complex polarization aberrations[39], suggests their potential for applications in nanoscale[40,41], optical communications[42], and photonic computing[43].

    The orbital angular momentum (OAM) of light plays a crucial role in constructing skyrmionic textures in both paraxial and nonparaxial beams[20,26,27], as the localized vortex within these topological quasiparticles arises from the twist and curl of the phase structure. Of the various skyrmionic objects proposed in optical systems, they are formed within the spatial domain, which can be regarded as 'longitudinal skyrmions' induced by the longitudinal OAM, where the skyrmion plane is perpendicular to the propagation direction, and the localized spin structures remain static over time [Fig. 1(a)]. This results in a helical twist within the skyrmion tube, attributed to the Gouy phase shift among propagation. Transverse OAM embedded within

pulsed wavepackets, on the other hand, provides an additional degree of freedom for manipulating light topology[44]. Driven by advancements in ultrafast lasers, spectral manipulation, and pulse characterization techniques, spatiotemporal sculpting of light has become feasible within time-dependent optical pulses[44–51]. While numerous topological structures have been observed in time-varying electromagnetic excitations, such as toroidal light pulses[52] and optical hopfions[53], these structures primarily reside within the spatial domain and their topological properties are limited to scalar fields. Vectorial topological quasiparticles in the spatiotemporal domain remain largely unexplored.

In this work, we theoretically propose and experimentally demonstrate optical spatiotemporal skyrmions in the space-time domain, achieved through the utilization of transverse OAM and vectorial shaping of pulsed light. By extending vectorial control from a purely spatial framework to a fully spatiotemporal paradigm, we construct the fundamental spatiotemporal skyrmion texture using spatiotemporal Gaussian mode pulses and spatiotemporal vortex mode pulses with orthogonal polarization states, which are carefully aligned both temporally and spatially [Fig. 1(b)]. The spatiotemporal skyrmion textures emerge within the vector field's spatiotemporal distribution, encompassing all possible polarization states across both the time and spatial dimensions. In these textures, the skyrmions exhibit no helical twisting perpendicular to the skyrmion plane ($x$-$t$ plane) due to the uniform vector sculpting along the y-axis, resulting in a perfect skyrmion tube with identical skyrmion slice. The helicity of spatiotemporal skyrmion gradually evolves upon propagation caused by the interplay between medium dispersion and spatial diffraction. Since transient spatiotemporal skyrmions maintain a stable topological texture in the spatiotemporal dimension, they hold significant potential for nontrivial light-matter interactions and offer additional degrees of freedom for information transfer.

## Theoretical analysis and numerical simulations

The optical pulse wavepackets can be expressed as the product of the carrier wave and an envelope function as $E(t,x,y,z) = \Psi(t,x,y,z)\exp(i\omega_0 t - ikz)$, where $\omega_0 = k_0 c = 2\pi c/\lambda_0$ is the central angular frequency and $k_0$ is the wavenumber in a vacuum. Within the framework of the scalar, paraxial, and narrow bandwidth approximations, the evolution of the envelope wavepacket in a uniform dispersive medium is governed by[50–52]

$$\frac{\partial^2 \Psi}{\partial x^2} + \frac{\partial^2 \Psi}{\partial y^2} - k_0 \beta_2 \frac{\partial^2 \Psi}{\partial \tau^2} + 2k_0 i \frac{\partial \Psi}{\partial z} = 0, \quad (1)$$

where $\Psi$ is the scalar envelope function, $\tau = t - z/v_g$ is the normalized local time, $z$ is the propagation distance, $v_g$ is the group velocity, $\beta_2$ is the group velocity dispersion (GVD) coefficient of the dispersive medium. In an anomalous GVD medium with $\beta_2 = -1/k_0$ (assuming the wavepacket is uniformly distributed along y-direction.), Equation (1) admits a closed-form solution as a spatiotemporal Laguerre-Gaussian (LG) wavepacket (at $z=0$), which could be expressed as[46,52]

$$\Psi_{\text{STLG}}(\tau,x) \propto \left(\frac{\sqrt{2(\gamma^2\tau^2+x^2)}}{w_0}\right)^{|l|} \exp\left(-\frac{\gamma^2\tau^2+x^2}{w_0^2}\right) L_p^{|l|}\left(\frac{2(\gamma^2\tau^2+x^2)}{w_0^2}\right) \exp\left[il\tan^{-1}\left(\frac{x}{\gamma\tau}\right)\right], \quad (2)$$

where $w_0$ is the waist radius; $L_p^{|l|}$ is the associated Laguerre polynomials of the order $p$ and $l$. $\gamma$ is a scaling factor of normalized local time.

Equation (2) describes the spatiotemporal LG mode as the scalar solution of Eq. (1). Maxwell's equations are inherently vectorial, describing the electric field as comprising three components: $E_x$, $E_y$, and $E_z$. For a monochromatic light propagating along the z-direction, and under the paraxial approximation, only the transverse components $E_x$ and $E_y$ are typically considered. These components define the spatially varying polarization state of light, known as vector light, a phenomenon extensively studied both theoretically and experimentally. However, the polarization dynamics of pulsed light, varying with both time and space, remain largely unexplored. Here, we introduce a wavepacket with spatiotemporally evolving polarization, whose Stokes parameters form a distinctive topological structure—a skyrmion—in the $t$-$x$ plane.

In monochromatic light, the spatial skyrmion topology can be formed by a pair of spatial LG modes with orthogonal polarization states[26,27,33] as $\mathbf{E}(x,y) \propto LG_{0,0}(x,y)\mathbf{e}_R + LG_{0,1}(x,y)\mathbf{e}_L$. In pulsed light, the spatiotemporal skyrmion topology pulse can be formed by a pair of spatiotemporal LG pulses with orthogonal polarization states, which have the following expression

$$\mathbf{E}(\tau,x,y) \propto \exp\left(-\frac{y^2}{y_0^2}\right)\left[\cos c_0 LG_{0,0}(\tau,x)\mathbf{e}_R + \sin c_0 e^{i\phi_\gamma} LG_{0,1}(\tau,x)\mathbf{e}_L\right], \tag{3}$$

where $\mathbf{e}_R$, $\mathbf{e}_L$ represent right-circular polarization (RCP) and left-circular polarization (LCP) components of transverse electric field, and $LG_{p,l}$ is the LG mode characterized by radial and azimuthal indices $p$, $l$. The global phase difference between the two orthogonal polarization components is denoted by $\phi_\gamma$, which controls the helicity textures of topological quasiparticle. $c_0$ is a constant that determines the amplitude ratio of the orthogonal modes. $y_0$ and $\tau_0$ are the spatial and temporal width of the wavepacket along the $y$ and $\tau$ directions, respectively.

There are two typical methods for generating vector beams. The first method involves controlling the spatial distribution of the polarization state by manipulating the amplitude ratio and phase difference between the LCP and RCP components[27]. The second method exploits the properties of waveplates, which can rotate the polarization ellipse within the Stokes space[33,54,55]. Consequently, spatially varying waveplates can modify the spatial distribution of the polarization state[49,56]. In our work, spatiotemporal polarization distribution is realized by employing the first method. A spatiotemporal vortex wavepacket of $LG_{0,1}$ mode [Fig. 2(d)] and a Gaussian wavepacket of $LG_{0,0}$ mode [Fig. 2(a)] are first produced, corresponding to LCP and RCP polarization, respectively. When the two wave packets are carefully coaxially superimposed in the space-time plane [Fig. 2(g)], the spatiotemporal polarization distribution is determined by the amplitude ratio and phase difference between the RCP and LCP components. The amplitude and phase distributions of their cross-sections of RCP and LCP ($y = 0$ plane) are shown in Figs. 2(b, c) and 2(e, f). At the center of the wave packet ($y = 0$ plane), the amplitude ratio of the Gaussian mode ($LG_{0,0}$) to the vortex mode ($LG_{0,1}$) is maximal, resulting in right-handed polarization. At the edges ($y = 0$ plane), this ratio is minimal, producing left-handed polarization. The intensity ratio decreases continuously from infinity at the center of the wavepacket to zero at its edge along the radial direction of $y = 0$ plane. At a specific radius where the Gaussian and vortex mode amplitudes are equal, linear polarization is observed. The helicity of skyrmion is controlled by the phase difference between the RCP and LCP components. Since the Gaussian mode has a constant phase, this phase difference is entirely determined by the vortex phase, which defines the orientation angle of the linear polarization. The spatiotemporal polarization distribution can be depicted using a polarization ellipse [Fig. 2(l)] or Stokes parameters [Figs. 2(j, k)]. The out-of-plane and in-plane components of the Stokes parameters are shown in Figs. 2(h, i), respectively. The out-of-plane component gradually transitions from 1 to −1 along the radial direction [Fig. 2(h)]. Meanwhile, the in-plane component, represented by [$\tan^{-1}(S_1/S_2)$], undergoes a full $2\pi$ rotation along the azimuthal direction [Fig. 2(i)]. The Stokes orientation distribution in $x$-$t$ plane clearly illustrate the skyrmion topological texture [Figs. 2(j, k)]. Through stereographic projection, the normalized reduced Stokes parameters in the spatiotemporal plane are mapped onto the Poincaré sphere, covering it exactly once and yielding a skyrmion number of 1.

## Experimental results

Utilizing the Fourier transform relationship between the space-time domain($t$, $x$) and the spatiotemporal frequency domain ($\Omega$, $\xi$), we construct corresponding spatial-spectrum to generate a spatiotemporal LG beam in space-time domain[46,52]. Accordingly, we experimentally synthesize the spatiotemporal skyrmion wavepackets by 2D spatial light modulation and ultrafast pulse shaping[44,52,57].

A chirped horizontal polarized pulsed beam is divided into an object pulse beam and a probe pulse beam by a polarization beamsplitter (BS1). The object pulse beam was polarized at 45° by HWP1, and then spatially spread along $y$-axis by a grating and collimated onto a spatial light modulator (SLM) through a cylindrical lens. The SLM plane can be regarded as the Fourier conjugate plane of the spatiotemporal plane. Iris1 filters the spectrum to maintain a Gaussian mode (see more details in the sec.1 of supplementary materials). The $x$-polarized component of object beam is modulated by a meticulously crafted hologram (see Methods section) to achieve the spatiotemporal $LG_{0,1}$ mode in 0th-order diffraction. The other diffraction orders are blocked by Iris2. In contrast, the residual $y$-polarized component is transmitted directly, maintaining a Gaussian shape. This arrangement ensures precise coaxial alignment between the $x$- and $y$-polarized components. The amplitude ratio and phase difference between them define the polarization state. In the LCP and RCP basis, a skyrmion topology is formed, while in the $x$- and $y$-polarization basis, a bimeron topology is realized. Here, the $x$- and $y$-polarizations are converted into RCP and LCP after passing through QWP2, forming a skyrmion topology. To fully characterize the spatiotemporal skyrmion wavepacket, we employ temporally sliced off-axis interference technique[58,51].

Briefly, the chirped probe pulse passes through a pulse compressor and a time delay line, then is de-chirped into a Fourier-transform-limited pulse (~120fs). The time-delayed probe pulsed beam interferes with the object pulsed beam, generating interference fringes that encode the spatial complex field of the wavepacket at a certain temporal slice. Scanning the probe pulse's time delay enables the reconstruction of the object wavepacket's spatiotemporal profile from the delay-dependent fringes[44]. HWP2 aligns the probe polarization with the object beam to enable interference. The spatiotemporal profiles of the RCP and LCP components are obtained through two separate 3D diagnostic measurement.

We experimentally generated the fundamental Neel-type spatiotemporal Stokes skyrmion topology wavepacket, as shown in Fig. 4. The intensity iso-surface of the RCP component exhibits a Gaussian mode, as it is derived from the $y$-polarized component, which cannot be modulated by the SLM, as shown in Fig. 4(a). The RCP component features a uniform sliced phase distribution [Fig. 4(b)] and a Gaussian-mode sliced amplitude profile [$y$=0 plane, Fig. 4(c)], though with slight deformations. The deformations typically caused by the laser spectrum and optical system imperfections. Figure 4(d) presents the reconstructed spatiotemporal 3D iso-intensity surface of LCP component, indicating a complicated spatiotemporally coupled 3D wavepacket, which resembles a hollow doughnut shaped structure. The sliced amplitude and phase distributions [Figs. 4(e, f)] of the LCP component clearly exhibit its spatiotemporal vortex characteristics, including the hollow intensity profile and helical phase structure, as shown in Fig. 4(e) and 4(f) respectively. A single-beam vector beam generation method, employing 0th-order diffraction complex amplitude holographic encoding, enables precise spatial coaxiality and temporal synchronization of the RCP and LCP components [Fig. 4(g)], yielding the designed polarization distribution depicted in Fig. 4(l). The out-of-plane component $S_3$ of the normalized Stokes parameters [Fig. 4(h)] exhibits a continuous variation from +1 at the core to -1 at the boundary, corresponding to the transition from the North Pole to the South Pole of the Poincaré sphere. Concurrently, the in-plane component $\tan^{-1}(S_1/S_2)$] [Fig. 4(i)] undergoes a phase evolution from 0 to $2\pi$, encompassing the entire azimuthal range of the Poincaré sphere. The three-dimensional normalized Stokes field in spatiotemporal space spans the entire Poincaré sphere, constructing a skyrmion topology [Figs. 4(j, k)]. Despite slight distortions in the experimental results, the topological invariant remains preserved, with a skyrmion number of 0.92, highlighting the robustness of skyrmion topology. Additional spatiotemporal skyrmion topologies can be realized within this setup by adjusting the complex amplitude of the LCP component, as demonstrated through numerical simulations provided in the Sec. 2 of supplementary materials.

## Discussions

In monochromatic waves, the skyrmion plane ($x$, $y$) and the propagation dimension ($z$) define a time invariant, three-dimensional topological structure which can be trivially extended along the propagation direction, forming a skyrmion tube as shown in the left column of Fig. 5. The expansion of the intensity profile and $s_3$ distributions during propagation results in a distance-dependent variation of the skyrmion number [Figs. 5(a1–c1)]. This variation is attributed to the longitudinal OAM along the $z$-dimension, which twists the iso-phase surface, inducing a Gouy phase shift that drives the helical twisting of the skyrmion tube [Fig. 5(d1–f1)]. In contrast, the intensity profile and $s_3$ distributions ($x$-$t$ plane) for spatiotemporal skyrmion do not spread along $y$ dimension due to the uniform vector sculpting applied consistently across $x$-$t$ planes, as shown in Fig. 5(a2, b2). The skyrmion number remains constant along the axis of skyrmion tube (Fig. 5(c2)). The transverse OAM also induces an invariant helicity [Figs. 5(d2-f2)], even as the spatial distribution of wavepacket change.

The interplay between medium dispersion and spatial diffraction results in an intriguing propagation property for the spatiotemporal skyrmion. In dispersion-matched media, the propagation of spatiotemporal skyrmion pulses is governed by a precise balance between medium dispersion and spatial diffraction, leading to identical complex amplitude variations. Due to the form-invariance of spatiotemporal LG modes in dispersion-matched media, the beam spot size varies with propagation, as shown in Figs. 6(a1-a3). The Gouy phase shift [ Fig. 6(b1-b3) and 6(c1-c3)] of skyrmion pulses causes a uniform variation in skyrmion helicity along the propagation distance $z$ [Fig. 6(d)], while the $y$-dimension remains unaffected. At $z$ = 0 plane, the skyrmion tube is of the Neel type, transitioning to an intermediate type as $z$ approaches $z_R$, and eventually becoming a Bloch type as $z$ to infinity. In addition, the propagation of spatiotemporal skyrmion in free space could causes deformations in the skyrmion topology (as detailed in sec.3 of supplementary materials).

## Conclusion

In conclusion, we have extended skyrmion topology from the spatial to the spatiotemporal domain and demonstrated the generation of spatiotemporal skyrmions within ultrafast pulse wavepackets. These spatiotemporal skyrmion pulses, identified as vector solutions to Maxwell's equations, exhibit a stable topological structure with polarization states evolving across both time and space. Experimentally, we construct spatiotemporal skyrmions by vectorially sculpting spatiotemporal wavepackets. Unlike conventional spatial skyrmions formed by longitudinal OAM, spatiotemporal skyrmions are induced by transverse OAM and exhibit no helical twisting perpendicular to the skyrmion plane, ensuring enhanced stability against deformation. Additionally, we show that the helicity of spatiotemporal skyrmions gradually evolves during propagation due to the interplay of medium dispersion and spatial diffraction. The ability to maintain a robust topological texture in the spatiotemporal domain opens new possibilities for structured light-matter interactions, optical information encoding, and ultrafast photonic applications. By advancing vectorial control to the spatiotemporal regime, this work broadens the scope of optical skyrmions and provides new opportunities for engineering topological states of light.

## Methods

### The definition and calculation of the Stokes parameters

The Stokes parameters, as a pseudovector, characterize the polarization state of light under paraxial conditions ($E_z = 0$) and can be represented on the Poincaré sphere, where polarization states are mapped to specific points. For normalized Stokes parameters ($s_1$, $s_2$, $s_3$), its direction generally aligns with the spatial coordinate axes ($x, y, z$). In the case of monochromatic light propagating along the $z$-axis, skyrmion topology can be constructed in the $x$-$y$ plane. The $s_3$ component corresponds to circular polarization states, with RCP aligned along $+z$ and LCP along $-z$. Consequently, $s_3$ is conventionally mapped to the $z$-axis, while $s_1$ and $s_2$ lie within the $x$-$y$ plane.

To extend this concept to a pulsed wavepacket, the skyrmion topology is constructed in the $t$-$x$ plane using the normalized reduced Stokes parameters ($s_1$, $s_2$, $s_3$) = ($S_1/S_0$, $S_2/S_0$, $S_3/S_0$). Here, $s_3$ is mapped to the axis perpendicular to the topological plane ($t, x$), while $s_1$ and $s_2$ is mapped along $t$- and $x$-axis respectively, providing an intuitive visualization of spatiotemporal skyrmion topology. The stokes parameter ($S_0$, $S_1$, $S_2$, $S_3$) can be calculated as follows:

$$\begin{aligned} S_0 &= |E_R|^2 + |E_L|^2, \\ S_1 &= 2\,\mathrm{Re}(E_R \cdot E_L^*), \\ S_2 &= -2\,\mathrm{Im}(E_R \cdot E_L^*), \\ S_3 &= |E_R|^2 - |E_L|^2. \end{aligned} \qquad (4)$$

Where the $E_R$ and $E_L$ represent the RCP and LCP component of optical pulse wavepacket. The topological property of an optical skyrmion is determined by the skyrmion number, which represents the number of times the Stokes parameters ($s_1$, $s_2$, $s_3$) wraps around a unit sphere, expressed as

$$N_s = \frac{1}{4\pi} \iint \mathbf{s} \cdot \left( \frac{\partial \mathbf{s}}{\partial \xi} \times \frac{\partial \mathbf{s}}{\partial \eta} \right) d\xi d\eta, \qquad (5)$$

where $\xi$ and $\eta$ denote the orthogonal axis of skyrmion plane.

### Parameters for spatiotemporal holographic pulse shaper

The laser source is a home-built Yb:fiber laser with an all-normal-dispersion (ANDi) configuration. The laser spectrum centered at 1030 nm with spectral bandwidth of 10 nm. At the output, the laser beam is expanded to a beam diameter of 2mm to accommodate the long propagation distance in this experiment.

Based on the Fourier transform relationship between the space-time domain ($t, x$) and the spatiotemporal frequency domain ($\Omega, \xi$), we construct the spectrum in the spatiotemporal frequency domain to generate a spatiotemporal LG beam in space-time domain ($t, x$). The spatiotemporal Fourier transform is achieved by the spatially diffractive and temporally dispersive propagation, and then, an spatiotemporal LG wavepacket is synthesized [46,52].

In the vectorial spatiotemporal pulse shaper, the input 45-degree polarized pulse beam is directed onto a reflective blazed grating (grating density of 1200 lines/mm, incident angle approximately 45°). The grating exhibits different diffraction efficiencies for *x*-polarized and x-polarized light, so the angle of HWP1 needs to be adjusted to modify the amplitude ratio between the *x*-polarized and *y*-polarized components. The first-order diffracted light is then collimated by a cylindrical lens (focal length 100 mm) onto the SLM screen (Model: Holoeye GAEA-2.1-NIR-069, 3840 × 2160 pixels, with a pixel pitch of 3.74 μm). The SLM modulates only the x-polarized component which can be defined by the following expression

$$CGH = \mathrm{mod}\{\arg(\psi) + f_x \xi \cdot \mathrm{asinc}(1-\psi) + \pi \mathrm{asinc}(\psi) + GDD \cdot \Omega^2, 2\pi\}, \qquad (6)$$

where, $f_x$ represents the frequency of a linear phase ramp, with its depth determined by the modulus of the on-demand mode. Consequently, the undesired spatial-spectral energy is diffracted away from the optical axis by this phase grating after propagating a short distance from the SLM. A more detailed performance of the zero-order complex amplitude modulation can be found in Refs. [45,46,59].

## Acknowledgements
We acknowledge financial support from the National Natural Science Foundation of China (12434012, 12204309, 12304367); Shanghai Rising-Star Program (22YF1415200, 23YF1415800); Shanghai Postdoctoral Excellence Program (2023533).

## Author contributions
H.T. and X.R.L. proposed the original idea and initiated this project. H.T. X.L. and J.Z. completed the theory and simulations. H.T. X.L. N.Z H.F. G.C Q.C and J.Z. performed the experiments and analyzed the data. H.T. X.L. N.Z H.F. X.R.L. and Q.Z. prepared the manuscript. X.R.L. and Q.Z. supervised the project. All authors contributed to the discussion and writing of the manuscript.

## Competing interests
The authors declare no competing interests.

## Data availability.
The data that support the findings of this study are available from the corresponding author upon reasonable request.

## References

1. Skyrme, T. H. R. A non-linear field theory. *Proc. R. Soc. A* **260**, 127–138 (1961).

2. Skyrme, T. H. R. A unified field theory of mesons and baryons. *Nuclear Phys. B* **31**, 556–569 (1962).

3. Manton, N. *Skyrmions: A Theory of Nuclei*. (World Scientific, New Jersey, 2022).

4. Al Khawaja, U. & Stoof, H. Skyrmions in a ferromagnetic Bose–Einstein condensate. *Nature* **411**, 918–920 (2001).

5. Ruostekoski, J. & Anglin, J. R. Creating Vortex Rings and Three-Dimensional Skyrmions in Bose-Einstein Condensates. *Phys. Rev. Lett.* **86**, 3934–3937 (2001).

6. Leslie, L. S., Hansen, A., Wright, K. C., Deutsch, B. M. & Bigelow, N. P. Creation and Detection of Skyrmions in a Bose-Einstein Condensate. *Phys. Rev. Lett.* **103**, 250401 (2009).

7. Foster, D. *et al.* Two-dimensional skyrmion bags in liquid crystals and ferromagnets. *Nat. Phys.* **15**, 655–659 (2019).

8. Duzgun, A. & Nisoli, C. Skyrmion Spin Ice in Liquid Crystals. *Phys. Rev. Lett.* **126**, 047801 (2021).

9. Wu, J.-S. & Smalyukh, I. I. Hopfions, heliknotons, skyrmions, torons and both abelian and nonabelian vortices in chiral liquid crystals. *Liq. Cryst. Rev.* **10**, 34–68 (2022).

10. Mühlbauer, S. *et al.* Skyrmion Lattice in a Chiral Magnet. *Science* **323**, 915–919 (2009).

11. Yu, X. Z. *et al.* Real-space observation of a two-dimensional skyrmion crystal. *Nature* **465**, 901–904 (2010).

12. Romming, N. *et al.* Writing and Deleting Single Magnetic Skyrmions. *Science* **341**, 636–639 (2013).

13. Maccariello, D. *et al.* Electrical detection of single magnetic skyrmions in metallic multilayers at room temperature. *Nat. Nanotechnol.* **13**, 233–237 (2018).

14. Fert, A., Cros, V. & Sampaio, J. Skyrmions on the track. *Nat. Nanotechnol.* **8**, 152–156 (2013).

15. Sampaio, J., Cros, V., Rohart, S., Thiaville, A. & Fert, A. Nucleation, stability and current-induced motion of isolated magnetic skyrmions in nanostructures. *Nat. Nanotechnol.* **8**, 839–844 (2013).

16. Nagaosa, N. & Tokura, Y. Topological properties and dynamics of magnetic skyrmions. *Nat. Nanotechnol.* **8**, 899–911 (2013).

17. Göbel, B., Mertig, I. & Tretiakov, O. A. Beyond skyrmions: Review and perspectives of alternative magnetic quasiparticles. *Phys. Rep.* **895**, 1–28 (2021).

18. Shen, Y. *et al.* Optical skyrmions and other topological quasiparticles of light. *Nat. Photonics* (2023).

19. Tsesses, S. *et al.* Optical skyrmion lattice in evanescent electromagnetic fields. *Science* **361**, 993–996 (2018).

20. Luping Du, Aiping Yang, Anatoly V. Zayats, & Xiaocong Yuan. Deep-subwavelength features of photonic skyrmions in a confined electromagnetic field with orbital angular momentum. *Nat. Phys.* **15**, 650–654 (2019).

21. Gutiérrez-Cuevas, R. & Pisanty, E. Optical polarization skyrmionic fields in free space. *J. Opt.* **23**, 024004 (2021).

22. Lei, X. *et al.* Skyrmionic spin textures in nonparaxial light. *Adv. Photon.* **7**, (2025).

23. Xinrui Lei *et al.* Photonic Spin Lattices: Symmetry Constraints for Skyrmion and Meron Topologies. *Phys. Rev. Lett.* **127**, 237403 (2021).

24. Zhang, Q. *et al.* Periodic dynamics of optical skyrmion lattices driven by symmetry. *Appl. Phys. Rev.* **11**, 011409 (2024).

25. Lin, M., Liu, Q., Duan, H., Du, L. & Yuan, X. Wavelength-tuned transformation between photonic skyrmion and meron spin textures. *Appl. Phys. Rev.* **11**, 021408 (2024).

26. Gao, S. *et al.* Paraxial skyrmionic beams. *Phys. Rev. A* **102**, 053513 (2020).

27. Shen, Y., Martínez, E. C. & Rosales-Guzmán, C. Generation of Optical Skyrmions with Tunable Topological Textures. *Acs Photonics* **9**, 296–303 (2022).

28. Cisowski, C., Ross, C. & Franke-Arnold, S. Building Paraxial Optical Skyrmions Using Rational Maps. *Adv. Photonics Res.* 2200350 (2023).

29. McWilliam, A. *et al.* Topological Approach of Characterizing Optical Skyrmions and Multi-Skyrmions. *Laser Photonics Rev.* 2300155 (2023).

30. Król, M. *et al.* Observation of second-order meron polarization textures in optical microcavities. *Optica* **8**, 255 (2021).

31. Lin, W., Ota, Y., Arakawa, Y. & Iwamoto, S. Microcavity-based generation of full Poincaré beams with arbitrary skyrmion numbers. *Phys. Rev. Res.* **3**, 023055 (2021).

32. Shen, Y. Topological bimeronic beams. *Opt. Lett.* **46**, 3737 (2021).

33. Teng, H., Zhong, J., Chen, J., Lei, X. & Zhan, Q. Physical conversion and superposition of optical skyrmion topologies. *Photon. Res.* **11**, 2042 (2023).

34. Singh, K., Ornelas, P., Dudley, A. & Forbes, A. Synthetic spin dynamics with Bessel-Gaussian optical skyrmions. *Opt. Express* **31**, 15289 (2023).

35. Wang, A. A. *et al.* Generalized Skyrmions. Preprint at https://doi.org/10.48550/arXiv.2409.17390 (2024).

36. Rao, A. S. Optical skyrmions in the Bessel profile. *J. Opt. Soc. Am. A* **41**, 1059–1069 (2024).

37. Ornelas, P., Nape, I., De Mello Koch, R. & Forbes, A. Non-local skyrmions as topologically resilient quantum entangled states of light. *Nat. Photonics* (2024).

38. Beckley, A. M., Brown, T. G. & Alonso, M. A. Full Poincaré beams. *Opt. Express* **18**, 10777 (2010).

39. Wang, A. A. *et al.* Topological protection of optical skyrmions through complex media. *Light-Sci. Appl.* **13**, 314 (2024).

40. Aiping Yang *et al.* Spin-Manipulated Photonic Skyrmion-Pair for Pico-Metric Displacement Sensing. *Adv. Sci.* **10**, 2205249 (2023).

41. Xinrui Lei, Luping Du, Xiaocong Yuan, & Anatoly V. Zayats. Optical spin–orbit coupling in the presence of magnetization: photonic skyrmion interaction with magnetic domains. *Nanophotonics* **10**, 3667–3675 (2021).

42. He, T. *et al.* Optical skyrmions from metafibers with subwavelength features. *Nat. Commun.* **15**, 10141 (2024).

43. Wang, A. A. *et al.* Unlocking new dimensions in photonic computing using optical Skyrmions. Preprint at https://doi.org/10.48550/arXiv.2407.16311 (2024).

44. Chong, A., Wan, C., Chen, J. & Zhan, Q. Generation of spatiotemporal optical vortices with controllable transverse orbital angular momentum. *Nat. Photonics* **14**, 350–354 (2020).

45. Cao, Q., Zhang, N., Chong, A. & Zhan, Q. Spatiotemporal hologram. *Nat. Commun.* **15**, 7821 (2024).

46. Liu, X. *et al.* Spatiotemporal optical vortices with controllable radial and azimuthal quantum numbers. *Nat. Commun.* **15**, 5435 (2024).

47. Liu, X., Liang, C., Cao, Q., Cai, Y. & Zhan, Q. Ultrafast bursts of tailored spatiotemporal vortex pulses. Preprint at https://doi.org/10.48550/arXiv.2407.19747 (2024).

48. Cao, Q. *et al.* Non-spreading Bessel spatiotemporal optical vortices. *Sci. Bull.* **67**, 133–140 (2022).

49. Cao, Q., Zheng, P. & Zhan, Q. Vectorial sculpturing of spatiotemporal wavepackets. *APL Photonics* **7**, (2022).

50. Zhan, Q. Spatiotemporal sculpturing of light: a tutorial. *Adv. Opt. Photonics* **16**, 163 (2024).

51. Liu, X., Cao, Q. & Zhan, Q. Spatiotemporal Optical Wavepackets: from Concepts to Applications. *Photonic Insights* **3**, R08 (2024).

52. Wan, C., Cao, Q., Chen, J., Chong, A. & Zhan, Q. Toroidal vortices of light. *Nat. Photonics* **16**, 519–522 (2022).

53. Wan, C., Shen, Y., Chong, A. & Zhan, Q. Scalar optical hopfions. *eLight* **2**, 22 (2022).

54. Simon, R. & Mukunda, N. Universal SU(2) gadget for polarization optics. *Phy. Lett. A* **138**, 474–480 (1989).

55. Sit, A., Giner, L., Karimi, E. & Lundeen, J. S. General lossless spatial polarization transformations. *J. Opt.* **19**, 094003 (2017).

56. Teng, H., Zhong, J., Chen, J., Lei, X. & Zhan, Q. Physical Conversion and Superposition of Optical Skyrmion Topologies. *Photon. Res.* (2023).

57. Kondakci, H. E. & Abouraddy, A. F. Diffraction-free space–time light sheets. *Nat. Photonics* **11**, 733–740 (2017).

58. Li, H., Bazarov, I. V., Dunham, B. M. & Wise, F. W. Three-dimensional laser pulse intensity diagnostic for photoinjectors. *Phys. Rev. ST Accel. Beams* **14**, 112802 (2011).

59. Davis, J. A., Cottrell, D. M., Campos, J., Yzuel, M. J. & Moreno, I. Encoding amplitude information onto phase-only filters. *Appl. Optics* **38**, 5004–5013 (1999).

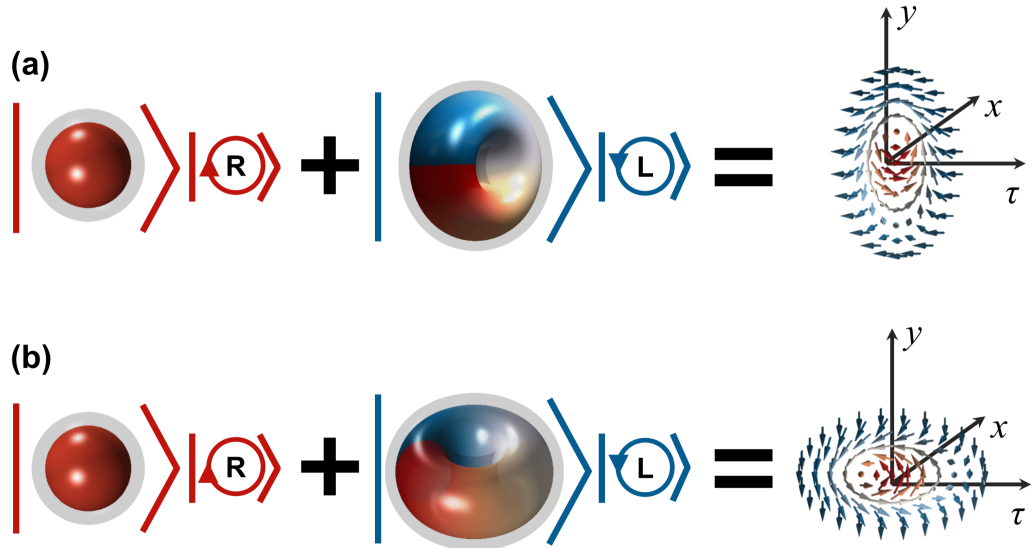

**Fig. 1** Spatial and spatiotemporal skyrmions. (a) Schematic of an optical spatial skyrmion constructed by a pair of spatial Laguerre–Gaussian (LG) modes with orthogonal polarization states. The normal vector of spatial skyrmion plane is parallel to the propagation direction. (b) Schematic of an optical spatiotemporal skyrmion wavepacket, constructed by a pair of spatiotemporal LG modes with orthogonal polarization states. In contrast to the former, the normal vector of spatiotemporal skyrmion plane is perpendicular to the propagation direction. The scalar spatiotemporal LG modes could be constructed by spatiotemporal hologram[45,46]. The colormap of the intensity iso-surfaces corresponds to the phases of LG modes, respectively.

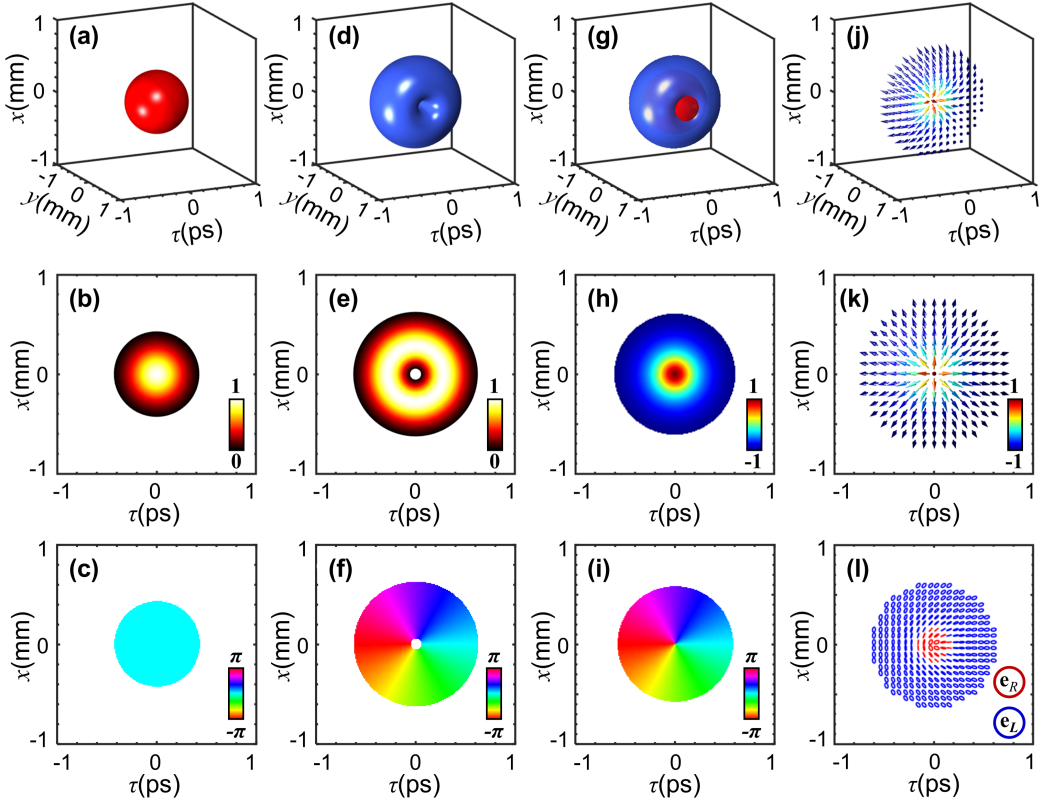

**Fig. 2** Simulation results of spatiotemporal skyrmion topology. (a) 3D iso-intensity (10% max) profiles of RCP spatiotemporal wavepacket with the corresponding (b) sliced intensity and (c) phase patterns at $y = 0$ plane. (d-f) 3D iso-intensity (10% max) profiles and corresponding sliced intensity and phase patterns (at $y = 0$ plane) of the LCP component. (g) A mixed 3D iso-intensity (10% max) profile of RCP and LCP spatiotemporal wavepacket. The normalized Stokes distribution of (h) out-of-plane ($S_3$) and (i) in-plane component [$\tan^{-1}(S_1/S_2)$]. (j) The 3D optical Stokes orientation distribution. (k) The top view of stokes orientation distribution. (l) The elliptical polarization distributions of spatiotemporal skyrmion topology. The definition and calculation of the Stokes parameters are detailed in the Methods section.

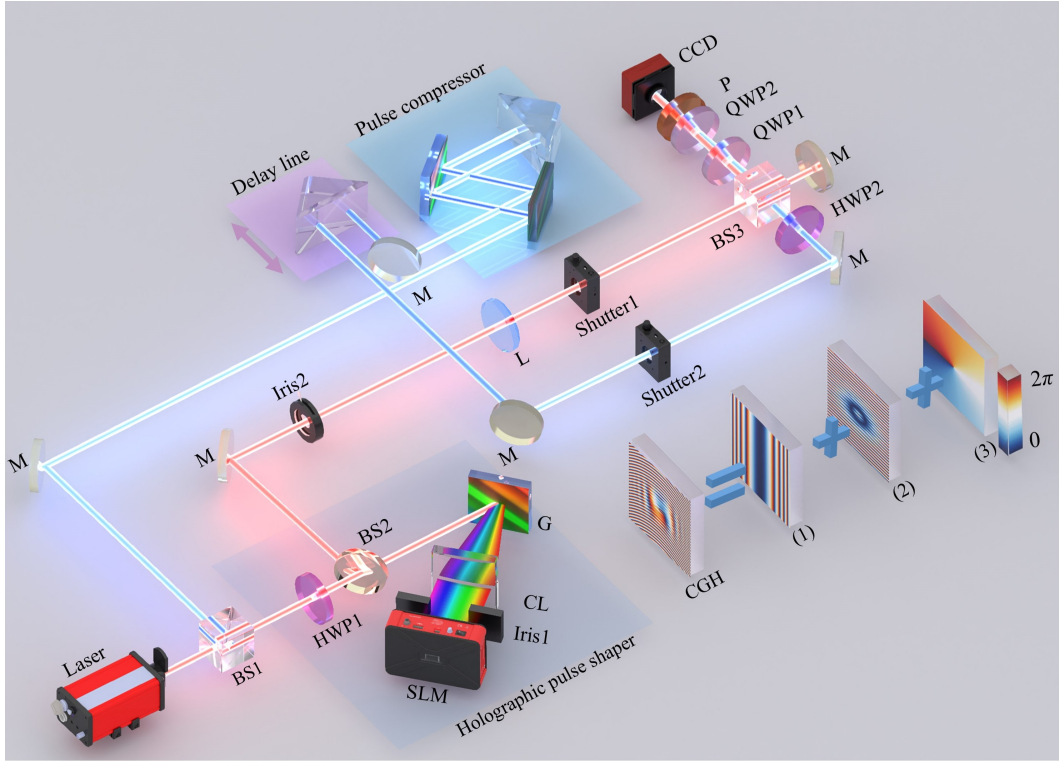

**Fig. 3** Optical setup for synthesizing and characterizing vectorial spatiotemporal skyrmion wavepackets. The apparatus comprises three components: (i) a 2D vectorial ultrafast holographic pulse shaper that incorporates a diffraction grating, a cylindrical lens, and an SLM, modulating only the *x*-polarization while leaving the *y*-polarization unaffected to control the amplitude ratio and phase difference between the *x*- and *y*-polarizations, thereby enabling polarization distribution control; (ii) a pulse compression system employing a parallel grating pair; and (iii) a time delay line system for fully reconstructing the 3D profile of the generated spatiotemporal wavepacket. Spatiotemporal holography used here must achieve accurate complex amplitude modulation in the zeroth order to ensure precise coaxial alignment between the *x*- and *y*-polarizations. The synthesized computer-generated hologram (CGH) enables complex-amplitude modulation of *x*-polarization, where the grating phase depth and delay regulate the amplitude and phase of the *x*-polarization component of incident pulsed beam, respectively[45,46,59]. The CGH embedded in the SLM consists of three components: (1) the phase distribution of a spatial-spectral $LG_{01}$ mode; (2) a GDD phase for managing the pre-chirp of the input pulse and balancing diffraction and dispersion; and (3) a phase-only diffraction grating with a phase depth modulated by the magnitude of Eq. (2) (see Methods). BS, beam splitter; HWP, half-wave plate; QWP, quarter-wave plate; P, polarizer; L, lens; M, mirror; G, grating; CL, cylindrical lens; SLM, spatial light modulator.

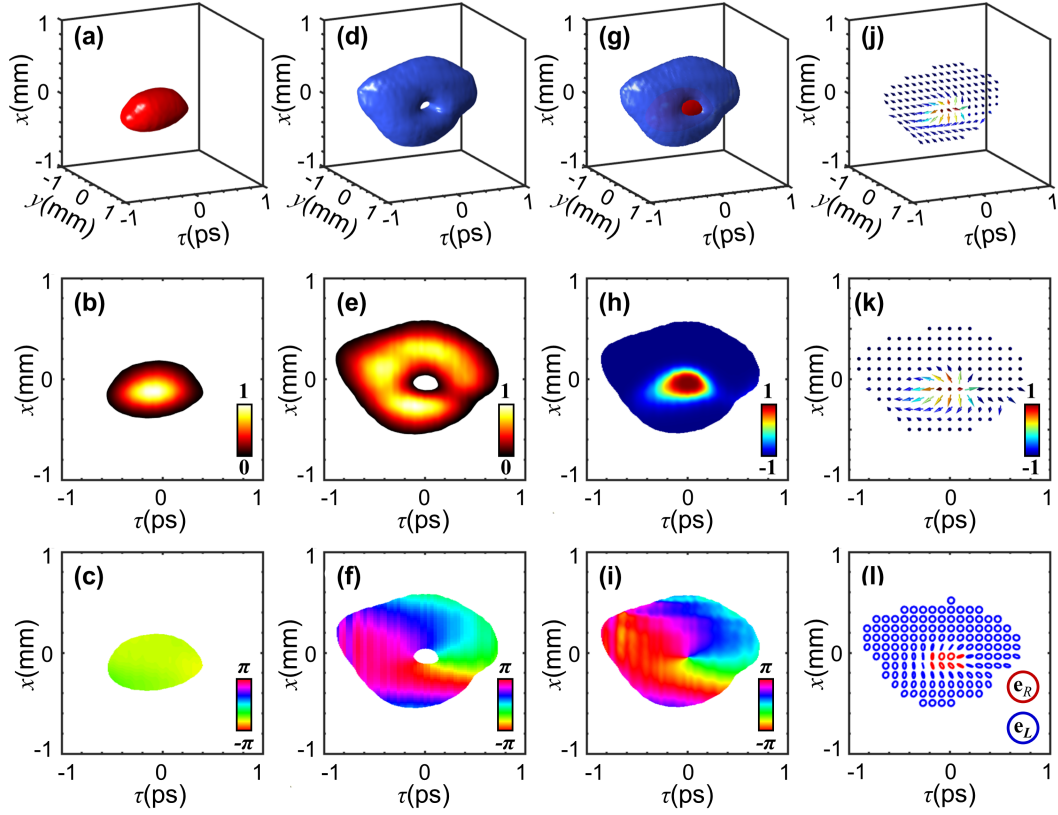

**Fig. 4** Experimentally synthesized spatiotemporal skyrmion wavepackets. (a-c) 3D iso-intensity (10% max) profiles and corresponding sliced intensity and phase patterns (at $y = 0$ plane) of the RCP component. (d–f) 3D iso-intensity (10% max) profiles and corresponding sliced intensity and phase patterns (at $y = 0$ plane) of the LCP component. (g) 3D iso-intensity (10% max) profiles of RCP and LCP components. The normalized Stokes distribution of (h) out-of-plane ($S_3$) and (i) in-plane component [$\tan^{-1}(S_1/S_2)$]. (j) The 3D optical Stokes orientation distribution. (k) The top view of Stokes orientation distribution. (l) The elliptical polarization distributions of spatiotemporal skyrmion.

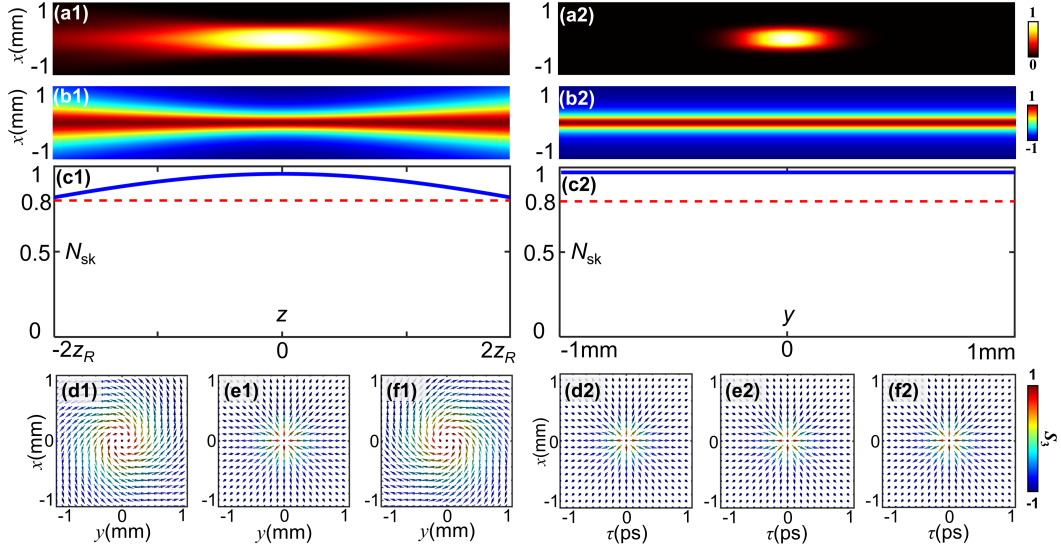

**Fig. 5** Skyrmion tubes of monochromatic light (a1–f1) and pulsed light (a2–f2). (a1, a2) Intensity distributions of the skyrmion tube for monochromatic light along the *z*-axis and pulsed light along the *y*-axis, respectively. (b1, b2) Out-of-plane Stokes orientation $s_3$ distributions of the monochromatic and pulsed skyrmion tubes, respectively. (c1, c2) Skyrmion number of the monochromatic and pulsed skyrmion tubes along the *z*- and *y*-axis, respectively. (d1–f1) Helicity variations of the monochromatic skyrmion tube along the *z*-axis, caused by the Gouy phase shift during propagation. (d2–f2) Helicity of the pulsed skyrmion tube along the *y*-axis, which remains nearly unchanged.

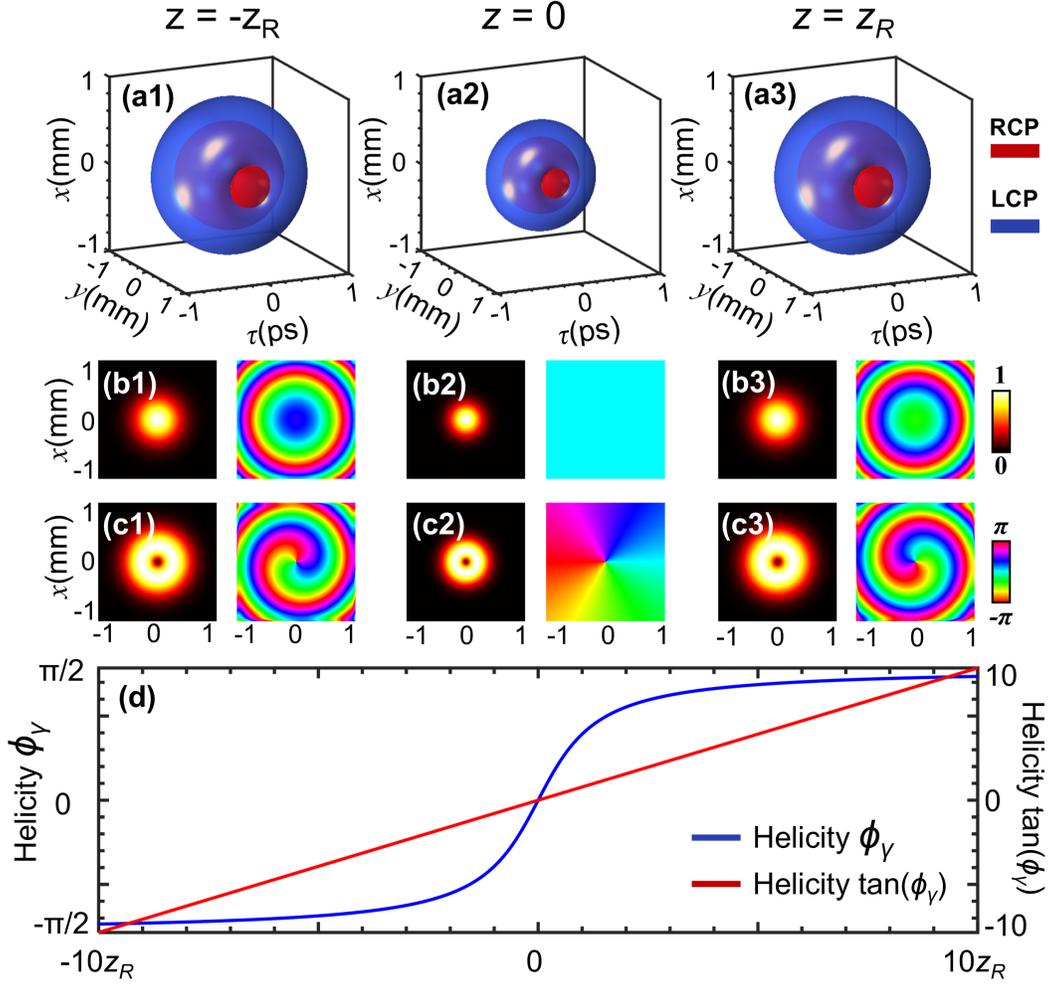

**Fig. 6** Numerical simulation of spatiotemporal skyrmion propagation in anomalous dispersive media. (a1–a3) The 3D iso-intensity (10% max) profiles of the RCP and LCP component of Spatialtemporal skyrmion at different propagation distances ($-z_R$, 0, and $z_R$). (b1–b3) and (c1–c3) The corresponding amplitude and phase distributions of RCP and LCP component. (d) Evolution of skyrmion helicity along the propagation dimension z.